\documentclass[prd,aps,floats,twocolumn]{revtex4}
\usepackage{amssymb}
\usepackage{amsmath}
\usepackage{mathrsfs}
\usepackage{latexsym}
\usepackage[dvips]{graphicx}
\begin{document}

\title{Constructing Dirac linear fermions in terms of non-linear Heisenberg spinors}

\

\author{M. Novello}
\affiliation{
Institute of Cosmology, Relativity and Astrophysics ICRA/CBPF \\
Rua Dr. Xavier Sigaud 150, Urca 22290-180 Rio de Janeiro,
RJ-Brazil}
\date{\today}

\begin{abstract}
We show that the massive (or massless) neutrinos can be described
as special states of Heisenberg nonlinear spinors.  As a
by-product of this decomposition a particularly attractive
consequence appears: the possibility of relating the existence of
only three species of mass-less neutrinos to such internal
non-linear structure. At the same time it allows the possibility
that neutrino oscillation can occurs even for massless neutrinos.
\end{abstract}

\vskip2pc
 \maketitle

There are evidences that neutrino changes from one flavor to another
as observed for instance in neutrino oscillations found by the
Super-Kamiokande Collaboration \cite{kamiokande}, \cite{Fukuda}.
This mix is understood as an evidence that the neutrino has a
 small mass \cite{Sigl}. This has important consequences not only in
local laboratory experiments but also in astrophysics and even in
cosmology. In a closely related path, the possibility that not only
left-handed but also right-handed neutrinos exist has recently
attracted interest, receiving a new treatment in a very imaginative
example presented in \cite{kapusta} dealing with the possibility of
neutrino superfluidity. The main idea requires the existence of an
interaction between neutrinos that in the case of small energy and
momentum can be described as a sort of Fermi process involving terms
like $(\overline{\nu}\nu) (\overline{\nu}\nu).$ If the $\nu$ 's are
the same, this interaction is nothing but an old theory of
Heisenberg concerning self-interacting fermions \cite{Heisenberg}.

Recent experiments \cite{Caso}  strongly support the idea that there
are only three neutrino flavours. Based on this and on the
possibility of mixing neutrino species it has been argued
\cite{Sigl}  that neutrino flavours $\nu_{\alpha}$ are combinations
of mass eigenstates $\nu_{i}$  of mass $m_{i}$ through a unitary $N
\times N$ matrix $U_{\alpha \, i}$.

It would be interesting  if we could describe all these properties
as consequences of the existence of a common root for the neutrino
species, e.g., if they are particular realizations of a unique
structure. In this paper we will develop a model of such idea and
work out a unified description of the three species of neutrinos by
showing that they can be considered as having a common origin on a
more fundamental nonlinear structure. Actually such property is not
exclusive for neutrinos but instead is typical for any Dirac fermion
(e.g., quark, electron). However as we shall see, the decomposition
of the Dirac fermion in terms of non-linear structure contains three
parameters ( associated to the Heisenberg self-interaction constant)
that separate different classes of Dirac spinors and three elements
for each class that could be associated to three types of particles
in each class. This form of decomposition may appear as if we were
inverting the common procedure and treating the simple linear case
of Dirac spinor as a particular state of a more involved
self-interacting nonlinear structure. This goes in the same
direction as some modern treatments in which linearity is understood
as a realization of a subjacent nonlinear structure. In this vein we
will examine the hypothesis that neutrinos are special states of
nonlinear Heisenberg spinors.

The argument is based on two fundamental equations: the linear Dirac
equation of motion
\begin{equation}
i\gamma^{\mu} \partial_{\mu} \, \Psi^{D} - M \, \Psi^{D} = 0
\label{domingo1}
\end{equation}
and the non linear Heisenberg theory \cite{sub}
\begin{equation}
i\gamma^{\mu} \partial_{\mu} \, \Psi^{H} - 2 s \, (A + iB
\gamma^{5}) \, \Psi^{H} = 0 \label{domingo2}
\end{equation}
in which the constant $s$ has the dimension of (lenght)$^2$ and
the quantities $A$ and $B$ are given in terms of the Heisenberg
spinor $\Psi^{H}$ as:
\begin{equation}
A \equiv \overline{\Psi}^{H} \Psi^{H} \label{domingo3}
\end{equation}
and
\begin{equation}
B \equiv i \overline{\Psi}^{H} \gamma^{5} \Psi^{H}
\label{domingo4}
\end{equation}

The Heisenberg spinor $\Psi^{H}$ can be depicted as a line making
$45$ degrees with each axis representing $\Psi_{L}$ and $\Psi_{R}$
in the two-dimensional plane $\pi$ generated by left-hand and
right-hand spinors as follows from the identity:
\begin{equation}
{\Psi}^{H} =  \Psi^{H}_{L} + \Psi^{H}_{R} = \frac{1}{2} (1 +
\gamma^{5})\Psi^{H} + \frac{1}{2} (1 - \gamma^{5})\Psi^{H}
\label{domingo5}
\end{equation}

The main outcome of the present paper is the proof of the statement
that massive or mass-less neutrino that satisfies Dirac equation
 (\ref{domingo1}) can be described as a deformation of the Heisenberg spinor
 in the $\pi$ plane. We are, indeed, claiming that it is
 possible to write the Dirac spinor as a
deformation of $\Psi^{H}$ in the plane $\pi$ given by
\begin{equation}
\Psi^{D} = e^{F} \, \Psi^{H}_{L} +  e^{G} \, \Psi^{H}_{R}
\label{domingo9}
\end{equation}
or, in other words, that the left and the right-handed Dirac spinor
are given by $\Psi^{D}_{L} = e^{F} \, \Psi^{H}_{L}$ and
$\Psi^{D}_{R} = e^{G} \, \Psi^{H}_{R}.$ What are the properties of
$F$ and $G$ in order that $\Psi^{D}$  satisfies equation
(\ref{domingo1})?

\subsection{The Inomata solution of Heisenberg dynamics}

In \cite{inomata} a particular class of solutions of Heisenberg
equation was set out. The interest on this class rests on the fact
that it directly  allows one to deal with the derivatives of the
spinor field, consequently allowing to obtain derivatives of any
associated function of the spinor. Let us briefly present this class
of spinors. The  analysis starts by the recognition that it is
possible to construct a sub-class of solution of Heisenberg dynamics
by imposing a restrictive condition given by
\begin{equation}
\partial_{\mu} \Psi = \left( a \, J_{\mu} + b\, I_{\mu}
\gamma^{5} \right)  \,\Psi \label{domingo6}
\end{equation}
where $a$ and $b$ are complex numbers of dimensionality
$(lenght)^{2}.$ This is a generalization, for the non-linear case,
of a similar condition in the linear case provided by plane waves,
that is, $
\partial_{\mu} \Psi = i k_{\mu} \, \Psi.$

A $\Psi$ that satisfies condition (\ref{domingo6}) will be called an
Inomata spinor. It is immediate to prove that if $\Psi$ satisfies
this condition it satisfies automatically Heisenberg equation of
motion if $a$ and $b$ are such that $ 2 s = i \, (a - b).$

 This is a rather strong condition that deals with simple
derivatives instead of the scalar structure obtained by the
contraction with $\gamma_{\mu}$ typical of Dirac or even for the
Heisenberg operators that appear in both equations (\ref{domingo1})
and (\ref{domingo2}). Prior to anything one has to examine the
compatibility of such condition which concerns all quantities that
can be constructed with such spinors. It is a remarkable result that
in order that the restrictive condition eq. (\ref{domingo6}) to be
integrable constants $a$ and $b$ must satisfy a unique constraint
given by $Re(a) - Re(b) = 0.$

Indeed, we have

$$ [\partial_{\mu}, \partial_{\nu} ] \,\Psi =  \left(a \, \partial_{[\mu} \,
J_{\nu]} +  b \, \partial_{[\mu} \, I_{\nu]} \, \gamma^{5} \right)
\, \Psi. $$ Now, the derivative of the currents yields
$$ \partial_{\mu} J_{\nu} -  \partial_{\nu} J_{\mu} = ( a + \overline{a})
 [J_{\mu}, \,J_{\nu}] + ( b + \overline{b}) \,
[I_{\mu} , \,  I_{\nu}], $$ and
$$ \partial_{\mu} I_{\nu} -  \partial_{\nu} I_{\mu} = ( a + \overline{a}
- b - \overline{b})  [J_{\mu}  \, I_{\nu} - I_{\mu} \, J_{\nu}].
$$
Thus the condition of integrability is given by
\begin{equation}
 Re (a) = Re (b).
 \label{20jan1735}
 \end{equation}

 It is a rather long and tedious work to show
that any combination $X$  constructed with $\Psi$ and for all
elements of the Clifford algebra, the compatibility condition
$[\partial_{\mu},
\partial_{\nu}] X = 0$ is automatically fulfilled once this unique
condition (\ref{20jan1735}) is satisfied.

 Let us now turn to some remarkable properties of I-spinors.

Lemma. The current four-vector $J^{\mu}$ is irrotational. The same
is valid for the axial-current $I^{\mu}$.

The proof that the vector $J^{\mu}$ is the gradient of a certain
scalar quantity is a simple direct consequence of its definition in
terms of H-spinors. However there is a further property that is
worth of mention: this scalar is nothing but the norm $J^{2}$ of the
current. Indeed, using equation (\ref{domingo6}), we have
\begin{equation}
\partial_{\mu} J_{\nu} = (a + \overline{a}) J_{\mu} J_{\nu} + (b +
\overline{b}) I_{\mu} I_{\nu} \protect\label{H13}
\end{equation}
This expression shows that the derivative of the four-vector current
is symmetric. Multiplying eq. (\ref{H13}) by $J^{\mu}$ it follows
then
\begin{equation}
J_{\mu} = \partial_{\mu} S \protect\label{H14}
\end{equation}
in which the scalar $S$ is written in terms of the norm $J^{2}
\equiv J_{\mu} J^{\mu}$:
\begin{equation}
S  = \frac{1}{a+ \overline{a}} \,  ln \sqrt{J^{2}}.
\protect\label{H15}
\end{equation}

Note that $ S = const. $ defines a  hypersurface in space-time such
that the current $J_{\mu}$ is not only geodesic but orthogonal to
$S.$ It follows
$$ \partial_{\mu} S \, \partial_{\nu} S \, \eta^{\mu\nu} = e^{2(a + \overline{a})
S},$$ or, defining the conformal metric $$g_{\mu\nu}^{c} \equiv e^{
2(a + \overline{a}) \, S} \, \eta_{\mu\nu} $$ we write
\begin{equation}
\partial_{\mu} S \, \partial_{\nu} S \, g^{\mu\nu}_{c} = 1,
\end{equation}
showing that $S$ is an eikonal in the associated conformal space.

Lemma. The two four-vectors $J_{\mu}$ and $I_{\mu}$ constitute a
basis for vectors constructed by the derivative $\partial_{\mu}$
operating on functionals of $\Psi$.

Proof. It is enough to show that this assertion is true for the
scalars $A$ and $B$. Indeed, we have:
\begin{equation}
\partial_{\mu} \, A =  (a + \overline{a}) \, A \, J_{\mu}  + (b -
\overline{b}) \, i B \,  I_{\mu} . \protect\label{H17}
\end{equation}
and
\begin{equation}
\partial_{\mu} \, B =  (a + \overline{a}) \, B \,  J_{\mu}  + (b
- \overline{b}) \, i A  I_{\mu} . \protect\label{H18}
\end{equation}

It then follows that the vector $I_{\mu}$ is a gradient too. Indeed,

\begin{equation}
\partial_{\mu} I_{\nu} = (a + \overline{a}) J_{\mu} I_{\nu} + (b + \overline{b}) J_{\nu}
I_{\mu} . \protect\label{H16}
\end{equation}

\begin{equation}
I_{\mu} = \partial_{\mu} R \protect\label{H152}
\end{equation}
in which the scalar $R$ is:
\begin{equation}
R = \frac{1}{b - \overline{b}} \, ln \left( \frac{ A - i
B}{\sqrt{J^{2}}} \right). \protect\label{H1532}
\end{equation}

\subsection{From Heisenberg to Dirac: How elementar is the neutrino?}

In this section we will describe an unexpected result of the Inomata
class ${\cal{IC}}$ which states that for any  spinor of
 ${\cal{IC}}$  it is possible to construct another spinor
 which satisfies the linear Dirac equation, the neutrino, for instance.
 In other words, we claim that a spinor that satisfies the linear
Dirac equation  may be constructed in terms of a non linear
structure. Let us prove this statement.  We start by defining the
deformation in the plane $\pi_{H}$ characterized by the left and
right-handed Heisenberg spinors by setting for the left and the
right-handed Dirac spinor the expressions
\begin{equation}
 \Psi^{D}_{L} = e^{F}  \, \Psi^{H}_{L}
 \end{equation}
 \begin{equation}
 \Psi^{D}_{R} = e^{G} \, \Psi^{H}_{R}
 \end{equation}

What are the properties of $F$ and $G$ in order that $\Psi^{D}$
satisfies Dirac equation? In order to answer this question we have
to make some additional calculations. From eq. (\ref{domingo6}) we
obtain
\begin{equation}
\partial_{\mu} \Psi^{H}_{L} = ( a \, J_{\mu} + b\, I_{\mu})
  \,\Psi^{H}_{L} \label{27jan17}
\end{equation}
\begin{equation}
\partial_{\mu} \Psi^{H}_{R} = ( a \, J_{\mu} - b\, I_{\mu}
)  \,\Psi^{H}_{R} \label{27jan171}
\end{equation}
Now comes the miracle that permits the accomplishment of our
procedure, which is the fact that the two vectors $J_{\mu}$ and
$I_{\mu}$ can be written as gradients of nonlinear expressions
under the form
\begin{eqnarray}
J_{\mu} &=& \partial_{\mu} S, \nonumber \\
I_{\mu} &=& \partial_{\mu} R, \label{quarta8}
\end{eqnarray}
 where $S$ and $R$ are given in eq. (\ref{H15}) and (\ref{H1532}).
From these equations it follows
\begin{equation}
\partial_{\mu} \Psi^{D}_{L} = \left( \frac{\partial F}{\partial S}  \, J_{\mu}
+ \frac{\partial F}{\partial R}  \, I_{\mu} \right)  \Psi^{D}_{L} +
( a \, J_{\mu} + b\, I_{\mu})  \,\Psi^{D}_{L}.
\end{equation}
and
\begin{equation}
\partial_{\mu} \Psi^{D}_{R} = \left( \frac{\partial G}{\partial S}  \, J_{\mu}
+ \frac{\partial G}{\partial R}  \, I_{\mu} \right)  \Psi^{D}_{R} +
( a \, J_{\mu} -  b\, I_{\mu})  \,\Psi^{D}_{R}.
\end{equation}

Multiplying these expressions by $ i \, \gamma^{\mu}$ it follows
that $ \Psi^{D} $ satisfies Dirac equation if $F$ and $G$ are given
by:
\begin{equation}
F = - \, \frac{1}{2} \, ( b - \overline{b} ) \, R + ( 2is -
\frac{1}{2} ( b - \overline{b} ) ) S + \frac{ i M} {a +
\overline{a}}
 \, e^{-(a + \overline{a}) S}
\end{equation}
\begin{equation}
G =  \, \frac{1}{2} \, ( b - \overline{b} ) \, R + ( 2 i s -
\frac{1}{2} ( b - \overline{b} )  ) S + \frac{ i M} {a +
\overline{a}}
 \, e^{-(a + \overline{a}) S}
\end{equation}
 To arrive at this result it is convenient to
use the formulas provided by Pauli-Kofink identities (see the
appendix) to obtain:
\begin{eqnarray}
J_{\mu} \, \gamma^{\mu} \, \Psi_{L} &=&  ( A  - i B ) \,
\Psi_{R} \nonumber \\
I_{\mu} \, \gamma^{\mu} \,  \Psi_{L} &=& -  ( A  - i
B) \, \Psi_{R} \nonumber \\
I_{\mu} \, \gamma^{\mu} \, \Psi_{R} &=&  ( A  + i B) \,
  \Psi_{L}  \nonumber \\
J_{\mu} \, \gamma^{\mu} \, \Psi_{R}  &=&  ( A  + i B ) \, \Psi_{L}.
\label{27dez915}
\end{eqnarray}
where
$$ A + i B = \frac{J^{2}}{ A - i B}. $$

 Thus, the linear Dirac field can be written in terms of the
non-linear Heisenberg field by:
\begin{equation}
 \Psi^{D}_{L} = \sqrt{\frac{J}{ A - i B}}  \,
  \exp{\left(\frac{i M}{(a +
\overline{a}) \,  J} + (2 i s -
  \frac{1}{2}( b - \overline{b} )  ) S
  \right)} \, \Psi^{H}_{L}
 \end{equation}
\begin{equation}
 \Psi^{D}_{R} = \sqrt{\frac{ A - i B}{J}}  \,
  \exp{\left(\frac{i M}{(a +
\overline{a}) \,  J} + (2 i s -
  \frac{1}{2}( b - \overline{b} )  ) S
  \right)}  \, \Psi^{H}_{R},
 \end{equation}
 where $ J \equiv \sqrt{J^{2}}.$
Using expression (\ref{H15}) we can simplify these expressions,
once we can write

$$ \exp{(2 i s -
  \frac{1}{2}( b - \overline{b} )  ) S } = J^{2 \sigma} $$
  where we have defined

  $$ \sigma \equiv \frac{i s - \frac{1}{4}( b - \overline{b})}{a +
  \overline{a}} = - \, \frac{i}{4} \, \frac{Im(a)}{Re(a)}.$$
 Then, finally, for the Dirac spinor
 \begin{equation}
\Psi^{D} =  \exp{\frac{i M}{(a + \overline{a}) \, J}} \, J^{2
\sigma} \,   \left( \sqrt{\frac{J}{ A - i B}}   \Psi^{H}_{L} +
\sqrt{\frac{ A - i B}{J}} \, \Psi^{H}_{R} \right) \label{7fev840}
 \end{equation}
 or, for the mass-less neutrino
\begin{equation}
\Psi^{D} =  J^{2 \sigma} \, \left( \sqrt{\frac{J}{ A - i B}}
\Psi^{H}_{L} + \sqrt{\frac{ A - i B}{J}} \, \Psi^{H}_{R} \right)
\label{7fev840}
 \end{equation}

This finally proves the following

Lemma: Free linear massive (or mass-less) Dirac field can be
represented as a combination of Inomata spinors satisfying the
non-linear Heisenberg equation.

We must analyze carefully the domain of parameters $a$ and $b$ once
the potentials $S$ and $R$ become singular in the imaginary axis and
in the real axis, respectively. Thus we can distinguish different
domains in the space of these two parameters. We set $a = a_{0} \,
e^{i\varphi}$ and $b = b_{0} \, e^{i \theta}.$ Then, the constraints
on these previously presented parameters, which allow for the
existence of the Inomata solution, are written under the form:
\begin{equation}
\frac{cos \varphi}{ \cos \theta} > 0,
\end{equation}
\begin{equation}
\cos \varphi \, ( tan \varphi -  tan \theta ) < 0, \label{14fev13}
\end{equation}
once the Heisenberg constant $s$ is positive. Let us name the
following sectors: $W_{1}$  for $ 0 < \varphi < \frac{\pi}{2};$
  $W_{2}$ for  $ \frac{\pi}{2} < \varphi < \pi;$  $W_{3}$  for $\pi < \varphi <
\frac{3 \pi}{2},$ and  $W_{4}$  for $ \frac{3 \pi}{2} < \varphi < 2
\pi.$ In an analogous way we define $Z_{1},$  $Z_{2},$  $Z_{3}$ and
$Z_{4}$ for similar sectors of $\theta.$ We distinguish then six
domains:
$$\Omega_{1} \equiv W_{1} \otimes Z_{1} $$
$$\Omega_{2} \equiv W_{4} \otimes Z_{1} $$
$$\Omega_{3} \equiv W_{4} \otimes Z_{4} $$
$$\Omega_{4} \equiv W_{2} \otimes Z_{2} $$
$$\Omega_{5} \equiv W_{3} \otimes Z_{2} $$
$$\Omega_{6} \equiv W_{3} \otimes Z_{3} $$
The missing domains $ W_{1}\otimes Z_{4}$ and $W_{2}\otimes Z_{3}$
are forbidden because they violate constraint (\ref{14fev13}). Thus,
for the massless case, equation (\ref{7fev840}) shows that different
choices of the parameters  $a$ and $b$ for a given value of constant
$s$  yield different spinor configurations $\Psi^{D}.$ This allows
us to write
\begin{equation}
\Psi^{D} = \sum_{\Omega_{i}}  \, c_{i} \,  \Gamma^{i,s}
\end{equation}
where $ \Gamma^{i,s} $ is defined by the rhs of equation
(\ref{7fev840}) and we have to sum over all possible independent
domains. Note furthermore that we are not obliged at this level to
specify the helicity. This expression exhibits the existence of a
degeneracy: for each Heisenberg theory characterized by a given
value of the self-coupling $s$ there are six distinct class of Dirac
spinors, which we could identify to three neutrinos and their
corresponding anti-neutrinos. A more careful analysis of the
topological sectors allows the identification of three pure states
of definite helicity, say $ \nu_{WZ}  = ( \nu_{11}, \nu_{44},
\nu_{41}) $ and their corresponding anti-states $ ( \nu_{22},
\nu_{33}, \nu_{32}).$ Note that the interacting neutrinos that
appears effectively in real processes are not in  general identical
to the pure cases. Thus in a very analogous way as is done in case
of massive eigen-states of neutrinos, we can write the interacting
ones $\nu_{\alpha}$ in terms of the above pure states $ \nu_{i}$ as
$$  \nu_{\alpha} = {\cal{M}}_{\alpha i} \, \nu_{i} $$
for $ \alpha$ and $i$ from 1 to 3; and their corresponding
conjugate anti-states. In this framework we can understand the
change of flavor of massless neutrinos.

Thus the above decomposition of Dirac fields in terms of Heisenberg
fields allows to understand neutrino oscillation without assuming
that such an oscillation should be a demonstration that neutrinos
are massive particles. The present mechanism of description allows
to understand oscillations occurring for mass-less neutrino.

Let us finally state that in addition, changing the value of $s$
allows the decomposition not only of neutrinos but also of other
fields in terms of fundamental Heisenberg spinors.

\subsection{Appendix}
\protect\label{algebraic}

The basis of the properties needed to analyze non-linear spinors
properties are contained in the Pauli-Kofink (PK) relation that
establishes a set of tensor relations concerning elements of the
four-dimensional Clifford $\gamma$-algebra. For any element $Q$ of
this algebra the PK relation states the validity of
\begin{equation}
(\bar{\Psi} Q \gamma_{\lambda} \Psi) \gamma^{\lambda} \Psi  =
(\bar{\Psi} Q \Psi)  \Psi  -  (\bar{\Psi} Q \gamma_{5} \Psi)
\gamma_{5} \Psi. \protect\label{H5}
\end{equation}
for $Q$ equal to $I$, $\gamma^{\mu}$, $\gamma_{5}$ and
$\gamma^{\mu} \gamma_{5}$. As a consequence of this relation we
obtain two extremely important consequences:
\begin{itemize}
 \item{The norm of currents $J_{\mu}$ and $I_{\mu}$ has the same
strength but opposite signs.}
 \item{The vectors  $J_{\mu}$ and $I_{\mu}$ are orthogonal.}
\end{itemize}

Indeed, using the PK relation we have
\begin{displaymath}
(\bar{\Psi}  \gamma_{\lambda} \Psi) \gamma^{\lambda} \Psi  =
(\bar{\Psi}  \Psi)  \Psi  -  (\bar{\Psi} \gamma_{5} \Psi)
\gamma_{5}
 \Psi.
\end{displaymath}
Multiplying by $\bar{\Psi}$ and using the definitions above it
follows
\begin{equation}
J^{\mu} J_{\mu} = A^{2} + B^{2}. \protect\label{H6}
\end{equation}
We also have
\begin{displaymath}
(\bar{\Psi} \gamma_{5} \gamma_{\lambda} \Psi) \gamma^{\lambda}
\Psi  = (\bar{\Psi} \gamma_{5} \Psi)  \Psi  - (\bar{\Psi} \Psi)
\gamma_{5} \Psi.
\end{displaymath}
From which it follows that the norm of $I_{\mu}$ is
\begin{equation}
I^{\mu} I_{\mu} = - A^{2} -  B^{2} \protect\label{H7}
\end{equation}
and that the four-vector currents are orthogonal
\begin{equation}
I_{\mu} J^{\mu} = 0. \protect\label{H71}
\end{equation}

\section*{Acknowledgements}
This work was partially supported by {\em Conselho Nacional de
Desenvolvimento Cient\'{\i}fico e Tecnol\'ogico} (CNPq) and {\em
Funda\c{c}\~ao de Amparo \`a Pesquisa do Estado de Rio de Janeiro}
(FAPERJ) of Brazil. I acknowledge the participants of the "Pequeno
Seminario" and in particular Drs J. M. Salim, I. Bediaga and N.
Pinto Neto for their comments on a previous version of this paper.


\begin{thebibliography}{100}
\bibitem{kamiokande} T. Kajita, Super-Kamiokande Coll., talk
presented at Neutrino '98 (434) 1998.
\bibitem{Fukuda} Y.Fukuda et al.Super-Kamiokande Coll., Phys.
Rev Lett 81, 1562 (1998)
\bibitem{Sigl} Gunter Sigl in Lectures on Astroparticle Physics
in the Procceedings of the XI Brazilian School of Cosmology and
Gravitation, Ed. M. Novello and S.E.P. Bergliaffa, AIP. See other
references therein.
\bibitem{kapusta} J. I. Kapusta, PRL 93, 251801 (2004).
\bibitem{Heisenberg} W. Heisenberg, Rev. Mod. Phys. 29, 269
(1957).
\bibitem{Caso} C. Caso et al. Particle Data Group, Eur. Phys. J. C
3, 1 (1998)
\bibitem{inomata} A. Inomata: in Proceedings of the International Seminar on
Relativity and Gravitation, Israel. Ed. C.G.Peres ans A.Peres,
Gordon and Breach, 1971.  See also A. Inomata and W.A.McKinley,
Phys. Rev. 140, B 1467 (1965) and M. Novello and R.C.Arcuri, Int. J.
Mod. Phys A 15, 2255 (2000).
\bibitem{sub} This equation represents a self-interacting
spinor field driven by a Lagrangian of the form typical of Fermi
processes, e.g., $L_{int} = s J_{\mu} J^{\mu}.$


\end{thebibliography}
\end{document}